\documentclass[11pt,twocolumn]{article}
\usepackage[sort,square,comma,numbers]{natbib}%
\usepackage{authblk}
\usepackage{bm}
\usepackage{graphicx}
\usepackage{dcolumn}
\usepackage{color}
\usepackage{float}
\usepackage{epstopdf}

\begin{document}

\title{
Impact of Screw and Edge Dislocation on the Thermal Conductivity of Nanowires and Bulk GaN
}

\author[1,2]{Konstantinos Termentzidis}

\author[1,3]{Mykola Isaiev}
\author[3]{Anastasiia Salnikova}
\author[4]{Imad Belabbas}
\author[2,1]{David Lacroix}
\author[5]{Joseph Kioseoglou}

\affil[1]{CNRS, LEMTA, UMR 7563, 54504 Vandoeuvre les Nancy, France}
\affil[2]{Universit$\acute{e}$ de Lorraine, LEMTA UMR 7563, 54504 Vandoeuvre les Nancy, France}
\affil[3]{Taras Shevchenko National University of Kyiv, 64/13, Volodymyrska Street, City of Kyiv, 01601, Ukraine}
\affil[4]{Laboratoire de Physico-Chimie des Mat\'{e}riaux et Catalyse, Un. Abderrahmane Mira, Beja\"{i}a (06000), Algeria}
\affil[5]{Department of Physics, Aristotle University of Thessaloniki, GR-54124 Thessaloniki, Greece}

\date{\today}
             
\maketitle

\begin{abstract}
We report on thermal transport properties of wurtzite
GaN in the presence of dislocations, by using molecular dynamics simulations. A variety of
isolated dislocations in a nanowire configuration were analyzed and found to reduce considerably the 
thermal conductivity while impacting its temperature dependence in a different manner. We
demonstrate that isolated screw dislocations reduce the thermal conductivity by a factor of two,
while the influence of edge dislocations is less pronounced. The relative reduction of
thermal conductivity is correlated with the strain energy of each of the five studied types of dislocations and the nature of the bonds around the dislocation core. The
temperature dependence of the thermal conductivity follows a physical law described by a
T$^{-1}$ variation in combination with an exponent factor which depends on the material's
nature, the type and the structural characteristics of the dislocation's core.
Furthermore, the impact of the dislocations density on the thermal conductivity of bulk GaN is examined. The variation and even the absolute values of the total thermal conductivity as a function of the dislocation density is similar for both types of dislocations. The thermal conductivity tensors along the parallel and perpendicular directions to the dislocation lines are analyzed.  The discrepancy of the anisotropy of the thermal conductivity grows in increasing the density of dislocations and it is more pronounced for the systems with edge dislocations.    
\end{abstract}

\section{Introduction}

Among III-V semiconductors, Gallium Nitride (GaN) has a plethora of applications namely in optoelectronics, piezoelectronics and high temperature electronic devices as it has a wide energy bandgap and strong chemical bonds, which makes it very stable at high temperatures\cite{gradecak05,huang02,jung12}. Another interesting property of GaN is its high thermal conductivity even for 1-D materials like: pristine or core/shell nanowires, nanorods, etc. Indeed, in GaN, the reduction of the thermal transport, when decreasing the dimensionality compared to the bulk, reaches a factor of $2$, while a factor of $10$ or more is observed in the case of Silicon\citep{Blandre2015,Termentzidis13a}. Nevertheless, as all III-V semiconductors, GaN exhibits a high density of structural defects, such as dislocations, which perturb the heat fluxes while creating overheated zones. This affects the performances and
reduces the life-time especially of electronic and optoelectronic devices. Threading dislocations are the major type of defects in heteroepitaxial, polar c-plane GaN layers~\cite{Alkauskas2016,Lozano2014}. Their effect on the electronic properties has been a matter of extensive investigation\cite{You2007,Veleschuk2013,Lymperakis2004}. 

The influence of threading dislocations on thermal transport has not been elucidated yet. However, the few existing theoretical studies  focus solely on the impact of screw dislocations~\cite{ni14,xiong14,Al-Ghalith2016}. Dislocations and compositional inhomogenities create energy states in the gap as well as band fluctuations. This leads to an increased nonuniformity which causes shunting, local heating and non-unique degradation~\cite{Veleschuk2013}. Dislocations are responsible for reverse breakdown and substantial leakage current in optoelectronics which reduce drastically the operating lifetime of devices. In order to control the behavior of the manufactured devices, we should be able to predict and tailor the impact of dislocations. Tuning the latter is a crucial issue for performing efficiency, stability and reliability of different electronic devices. From a fundamental point of view, the interaction of phonons with dislocations is far from being understood as only macroscopic models are suggested in the literature.

Phonons can be scattered by dislocations according to two distinguished mechanisms: a short-range and a long range one~\cite{Mamand2012}. In the first mechanism, phonons are scattered, in the immediate vicinity of the dislocation line, by the atoms located at the core, while in the second one scattering is due to the long ranged elastic strain field of dislocations. Theoretical models describing phonon-dislocation interactions can be categorized into static or dynamic~\cite{Zhu1991}. This refers to the position of the dislocation relative to the crystalline lattice during the phonon-dislocation interaction. Within molecular dynamics, both short and long-range scattering mechanisms and static or dynamic phonon-dislocation interactions are meant to be taken into account. 


The few experimental and theoretical studies considering the effect of the dislocation density on the thermal conductivity of bulk GaN revealed two regimes; in low density dislocation regime there is a plateau and thus the thermal conductivity is invariant as a function of the dislocation density, while in high density dislocation regime the thermal conductivity is severely degraded with a logarithmic dependence on the density \cite{Mion2006,Mion2016b,kotchetkov01,Zou2002}. At low temperature, the origin of the ``anomalous" behavior of the thermal conductivity in crystals with high dislocation density was assumed by Kogure and Hiki\cite{Kogure1975} to be related to quasi-local phonon modes, which are spatially restricted inside a narrow region around dislocations.

Recently, Kim et al~\cite{Kim109} proposed a methodology to introduce grain boundary dislocations in bulk bismuth telluride in order to increase the thermoelectric efficiency. These dislocations do not degradate the electrical conductivity but they reduce drastically the thermal conductivity. Bathula et al~\cite{Bathula2015} related the substantial enhancement of the thermoelectric figure of merit (ZT) of nanostructured Si$_{80}$Ge$_{20}$ alloy to an ultralow thermal conductivity, due to scattering of low-to-high wavelength heat-carrying phonons to point defects and dislocations. For the same alloy, Basu et al~\cite{Basu2014} found a ZT even higher than in the latter report and again dislocations were one of the explanations suggested for this enhancement. Murphy et al~\cite{Murphy2014} proposed to tailor thermal transport by controlling the defects and the phonon scattering times. They claimed that vacancies, impurities and dislocations affect the phonon transport through interatomic bonding forces, strain gradients around the defects and local changes in mass. By combining Monte Carlo and first principal calculations, Ma et al~\cite{Ma2013}, found that point defects and dislocations reduce the thermal conductivity of GaN mainly by restraining the transverse modes, while they have a little influence on the longitudinal ones. 

In SiC nanowires, Ni et al~\cite{ni14} showed that the core of a screw dislocation is a source of anharmonic phonon-phonon scattering and thus reduces the relaxation time of longitudinal acoustic phonons. Xiong et al~\cite{xiong14} studied the impact of one type of screw dislocation while changing the Burger vector from $1 \times a_0$ to $3 \times a_0$ in silicon nanowires and nanotubes. Very recently, Al-Ghalith-Ni-Dumitrica~\cite{Al-Ghalith2016}, studied PbSe and SiGe nanowires containing screw dislocations with various types of Burgers vectors. In the three previous studies, phonon scattering by dislocations has been suggested as an explanation for the thermal conductivity reduction. Gibbons et al~\cite{gibbons11} claimed that, in covalent crystals, impurities introduce localized vibrational modes which may absorb energy. These spatially localized vibrational modes can trap not only electric charges~\cite{kang14} but also phonons with vibrational lifetimes several hundred periods of oscillation~\cite{estreicher15,kang14}. These trapped phonons finally decay into lower frequency modes, thus reducing the thermal conductivity of the material~\cite{kang14}. Previous study of D. Spiteri et al~\cite{Spiteri2013}, has concluded that screw and edge dislocations lead to a reduction of about 39\% and 51\% of the bulk GaN thermal conductivity respectively.

The purpose of the present work is to study in a systematic way and to give physical insights on the influence of edge and screw dislocations, temperature and the density of dislocations on the thermal conductivity of GaN. In order to study the influence of isolated dislocations on thermal conductivity, first the structural configurations of nanowires are considered. Overall five different core configurations are modeled. This choice is related to the fact that it is not possible to obtain an isolated dislocation in a bulk system modeled with MD and with periodic boundary conditions. To ensure neutralization while adopting periodic boundary conditions (PBCs), one have to model minimum two opposite signed dislocations to achieve elimination of the total Burgers vector in the supercell. Thus, the presently used nanowire configurations offer the ability to investigate the influence of an isolated single dislocation on thermal conductivity. We have considered dislocations that extend along the [0001] direction which is the NWs’ growth direction. We then determine the influence of both: (i) the type of dislocation and (ii) the nature of the chemical bonds involved in its core, on the thermal conductivity along the dislocation line. For all five dislocation configurations, the temperature dependence of the thermal conductivity is calculated and a correction of the relation between thermal conductivity and temperature is proposed, with the introduction of an exponential factor. Finally, the effect of the dislocation density on the thermal conductivity is studied for a bulk system. The thermal anisotropy parallel and perpendicular to the dislocation lines is analyzed.    

This paper is organized as follows: in section II, model systems and the equilibrium molecular dynamics method are described, in section III the results on the thermal conductivity first of bulk, pristine GaN nanowires and of GaN nanowires with edge and screw dislocations are presented at $300$~K. Then, an analytically calculated strain energy using elasticity theory for both types of dislocations and the thermal conductivity temperature dependence is appraised. The effect of the dislocations' density on bulk GaN is also evaluated, revealing differences on the evolution of the thermal conductivity vectors parallel and perpendicular to the dislocation. Finally in section IV the conclusions of this work are reported.

section{Modelling defected nanowires and bulk}



In wurtzite GaN grown along the so-called polar direction, threading edge and screw dislocations have their lines laying in the [0001] direction. These dislocations are perfect and their Burgers vectors are respectively \textbf{b}$_e$ = 1/3[11$\overline{2}$0] and \textbf{b}$_s$=[0001]. For the edge dislocation, the Burgers vector has a magnitude of \textbf{b} = 3.189~\AA ~and is perpendicular to the line direction, while that of the screw one is parallel to the line direction having a magnitude of \textbf{b} = 5.214~\AA.

Linear elasticity theory~\cite{hirth82} provides a good framework to describe most of the properties of dislocations. Unless it fails to describe the core region, where the atomistic nature of the material is prevalent, elasticity theory is able to account for the displacement field of dislocations outside the core. Hence, in Cartesian orthogonal coordinate system where dislocations are assumed to be along the z direction, the displacement field of an edge dislocation for which the Burgers vector is assumed along the x is given by~\cite{hirth82}:

\begin{equation}
u_x = \frac{\textbf{b}}{2\pi} [arctan \frac{y}{x} + \frac{xy}{2(1-\nu)(x^2+y^2)}]
\end{equation}

\begin{equation}
u_y = - \frac{\textbf{b}}{2\pi} [ \frac{1-2\nu}{4(1-\nu)} ln(x^2 + y^2) +  \frac{x^2 - y^2}{4(1-\nu)(x^2+y^2)} ]
\end{equation}

Where $\nu$ represents the Poisson ratio. The displacement field of a screw dislocation along the z direction is given by~\cite{hirth82}:
 
\begin{equation}
u_z = \frac{\textbf{b}}{2\pi} arctan \frac{y}{x} 
\end{equation}

\begin{figure}[h!]
\begin{center}
  \includegraphics[width=6cm]{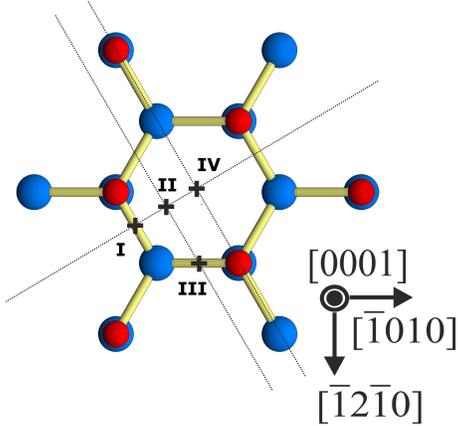}
  \caption{The hexagonal atomic unit of a wurtzite structure projected along the [0001] direction is presented. Four points I, II, III and VI, are denoted, which are used as origin of the displacement fields of the screw and edge dislocations. Large (blue) spheres denote Ga atoms, while small (red) spheres are N atoms.}
  \label{fig_1}
\end{center}
\end{figure}

In the present work, edge and screw dislocations were atomistically modeled by using the so-called: “supercell-cluster” hybrids~\cite{Blumenau2003}, within a multi-step procedure. A supercell-cluster hybrid contains a single dislocation and is periodic along the dislocation line direction, i.e. [0001]. Dislocations are introduced in a perfect wurtzite GaN crystal by applying the displacement field given by elasticity theory [Eq 1-3] which leads to reasonable initial atomic positions. Multiple atomic core configurations are obtained for either edge or screw dislocations, by considering the center of the aforementioned displacement fields is considered in various positions~\citep{bere02}. In figure~\ref{fig_1} the hexagonal atomic unit of a wurtzite structure, projected along the [0001] direction, is presented. It is proved that, concerning the edge dislocation: the 4-atom ring ($4E$), 5/7-atom ring ($5/7E$) and 8-atom ring ($8E$) core atomic configurations are formed when the origin of the displacement field at the positions I, II and III respectively~\cite{kioseoglou09,kioseoglou13,Belabbas2014}. Similarly for the screw dislocation, the single 6-atom ring ($S6S$) configuration is formed when the origin of the displacement field is positioned on the IV point, while when the origin is located on the I point, the double 6-atom ring configuration ($D6S$) is constructed~\cite{belabbas12,Belabbas2015a}.

Initially, very large supercell-cluster hybrids of about $30 000$ atoms containing an edge or a screw dislocation were constructed. These were pre-relaxed using empirical potentials following the methodology described by Kioseoglou et al~\cite{kioseoglou09,kioseoglou13} and Belabbas et al~\cite{belabbas12}. In a subsequent step smaller models of about 200 atoms were treated at the Density Functional Theory (DFT) level~\cite{Hohenberg64,Kohn1965}. These models were constructed from the previous ones and their lateral surfaces were passivated by fractionally charged pseudohydrogens. The passivation is implemented following the electron counting rule (ECR)~\cite{pashley89}.  Hence, the N dangling bond (DB) are saturated by pseudohydrogens that provide 3/4e and the Ga DBs by pseudohydrogens with a fractional charge of 5/4e. 

Our DFT calculations are performed using the ABINIT code~\citep{Gonze2009} within the local density approximation implementing modified pseudopotentials (MPP-LDA)~\cite{segev07}. The atomistic models were allowed to relax until the energy difference between successive relaxation steps was smaller than $2 \times 10^{-5}$~eV. A $50$ Ry cutoff was used for the plane-wave basis set while explicitly including Ga-3d electrons in the valence in order to improve the accuracy~\cite{Northrup1996}. The Brillouin zone was sampled through a 1 $\times$ 1 $\times$ 4 Monkhorst-Pack k-point grid.

The equilibrium positions were obtained through a relaxation based on the BFGS algorithm~\cite{Fischer92}. This procedure was conducted as follow~\cite{Kioseoglou2011}: for each cluster an outer region, consisting of the pseudohydrogens and the Ga, N atoms bonded to them, and an inner region which includes the rest of the atoms are considered. Initially the outer region is relaxed while the rest atoms were fixed. Afterwards, all atoms in the outer region are kept fixed to the positions derived from the previous step and the inner area atoms are allowed to fully relax. Periodic boundary conditions were implemented along the dislocation line direction, while fixed boundaries with a vacuum of $12$~\AA ~are considered along the two perpendicular directions. Following this procedure we circumvent the strain underestimation induced by implementation of the supercell-cluster hybrid approach\cite{Blumenau2003}. 


For the edge dislocation we modeled the $4E$, $5/7E$ and the $8E$, while for the screw dislocation, we considered the $S6S$ and the $D6S$ core configurations. The previous numbers correspond to the numbers of atoms forming the closest ring surrounding the dislocation line: 4, 5 and 7, 8 for the edge dislocation and a single six or a double six rings for the screw dislocation. In contrast to the report of Ni et al study\cite{ni14} in SiC, where the authors limited themselves to the configuration $S6S$, we have in addition considered the configuration $D6S$. The latter contains some homoelemental ``wrong" bonds (Ga-Ga, N-N), which being weaker than the Ga-N covalent bonds and they substantially influence the degree of the thermal conductivity reduction, as we will see in the following. All modeled nanowires have square cross-section with dimensions $4$~nm $\times$ $4$~nm and length of $10$~nm. The dislocations density with the specific cross-section is $70.9 \times 10^{11}$~cm$^{-2}$.

Concerning bulk systems and the study of the impact of the dislocation density on the thermal transport, as it was indicated in the introduction in order to ensure neutralization of the dislocations while applying periodic boundary conditions, four dislocations are created in the super-cells. The dislocations are constructed in pairs of opposite signs dipoles either along the [$\overline{1}$010] and the [$\overline{1}$2$\overline{1}$0] directions. We varied the dimensions of the cell, meaning also the distance of the dislocation pairs to achieve a variety of dislocation's density. It should be noticed that due to extremely high computational cost, we achieve to investigate super-cells up to about $10^{12}$~cm$^{-2}$ density of dislocations by the use of super-cells having about 370.000 atoms. This order of density of dislocations reaches the experimentally observed limits and it is sufficient to interpret the influence of dislocations in the real world GaN thin films and bulk materials. Another interesting point is the fact that the use of pairs of dislocations of opposite signs in MD simulations it is expected to lead to annihilation of the dislocations. Indeed, even at 300 K in all the super-cells having edge dislocations, annihilation is observed rapidly leading to the formation of perfect crystalline bulk GaN. In super-cells containing screw dislocations annihilation is observed for dislocation densities up to $4 \times 10^{13}$~cm$^{-2}$. In lower dislocation densities the attraction between the screw dislocations of opposite signs is not sufficient in order to annihilate them and hence we are able to investigate them. In order to avoid the annihilation of the edge dislocations, we implement a specific procedure in the LAMMPS code. We set a sufficient spring force independently to specific dislocation core atoms along the [$\overline{1}$010] direction and we preserved the edge dislocations in all the MD studies. However, those spring forces substantially influenced the phonon heat transfer and consequently the thermal properties along the [$\overline{1}$010] direction on supercells with edge dislocations are not reliable and are not considered in the present study.

In order to determine the thermal conductivity, we ran equilibrium molecular dynamics (EMD) simulations, using the LAMMPS~\cite{plimpton95,plimpton97}, a molecular dynamics code. The main idea of a EMD simulation is that the decay of the heat flux fluctuations is proportional to the thermal conductivity via the Green-Kubo formula~\cite{evans86}. As the described previously, the initial structures were created by ab-initio simulations and then they were replicated by $20$ times along the [0001] direction, while removing the terminating pseudo‐hydrogen atoms, to form a nanowire
with a free surface. The lattice parameters and the atomic positions of the model are rescaled to the equilibrium lattice parameters of the modified Stillinger-Weber potentential (MSWp)\cite{stillinger85}.

The timestep was set to $0.5$~fs, and then we obtained thermalisation with the NVT ensemble at $300$~K for $200$~ps, equilibration with NVE for $2$~ns and then averaging for $4-5$~ns. Periodic boundary conditions in all directions and a modified Stillinger-Weber interatomic potential\cite{stillinger85} are used. This potential reproduces well the thermal conductivity of the bulk GaN compared to the literature ($160$~W/mK). The simulation box was set 6~nm$\times$6~nm$\times$10~nm with $1$~nm empty space between the nanowire and the box, to ensure that there are no interactions between two adjacent nanowires. The thermal conductivity along the c-direction was computed using the Green-Kubo method which is based on the fluctuation-dissipation theorem:

\begin{equation}
\kappa_z = \frac{V}{k_BT^2} \int_0^{+\infty} \langle j_z(t)j_z(0) \rangle
\end{equation}  

where $V$ is the system volume, $k_B$ the Boltzmann constant and $j_z(t)$ is the instanteneous microscopic heat flux in the z-direction. We performed EMD simulations using 20 seeds of initial velocity to obtain results with satisfactory uncertainty (less than $10\%$).

\section{Results}

As wurtzite GaN has a hexagonal structure, the a-b and c directions, corresponding respectively to $[11\overline{2}0]$, $[10\overline{1}0]$ and $[0001]$, are not equivalent. The average thermal conductivity is found to be $148 \pm 8$~W/mK. The [0001] direction has slightly higher thermal conductivity tensor ($160 \pm 10$~W/mK) than the a and b-directions ($142 \pm 6$~W/mK). Here we found a thermal anisotropy of $11\%$, that is more pronounced than the $1\%$ suggested in the work of Slack et al\cite{Slack02}. In the latter, they explained this anisotropy due to the sound velocity anisotropy ($u_{\parallel} = 5.51 \times 10^5$~cm/s compared to $u_{\perp} = 5.56 \times 10^5$~cm/s) \cite{polian96}. The different orientations are easily to be explored with the EMD method as we can use the autocorrelation function of different heat flux directions. By integrating scattering terms that account for the effect of screw and edge dislocations, in a kinetic theory based model, Kotchetkov et al claimed that the large dispersion of the measured thermal conductivity of bulk GaN ($130-250$~W/mK) is due to the density of dislocations~\cite{kotchetkov01}. In the literature, there are several studies concerning the bulk thermal conductivity of GaN with several methods. By using the Non-Equilibrium Molecular Dynamics method (NEMD), Zhou et al found a value of $185$~W/mK~\cite{zhou09} while with the same method Liang et al~\cite{liang15} estimated the thermal conductivity in the [0001] direction equal to $166 \pm 11$~W/mK. They~\cite{liang15} explained the large discrepancy of their results obtained by different methods to the important contributions from long-mean free path phonons to the thermal conductivity. Kawamura et al calculated the bulk thermal conductivity with the EMD method and they found that at $300$~K, it ranges between $310-380$~W/mK~\cite{Kawamura2005}, while in their article they have included a figure with several values taken by the literature of the thermal conductivity of GaN, with the most of the values in the range of $150-200$~W/mK. 

Our purpose is not to make corrections of the interatomic potential or evaluate the impact of different methods on the thermal conductivity. We want to evaluate the impact of different types of dislocations and thus we consider that the calculated bulk thermal conductivity in our study is a good base to continue this investigation.

All modeled nanowires, pristine or with dislocations, are extended along the [0001] direction. A square shape cross-section of nanowires has been chosen to include the influence of the two main families of facets observed on wurtzite structure nanowires on the results: the ${10\overline{1}0}$ and the ${11\overline{2}0}$.  

The thermal conductivity of the pristine GaN nanowire has been calculated to be equal to 49 $\pm$3 W/mK.

This value is close to a previously reported one by Wang et al\citep{wang07}, who found by using the non-equilibrium molecular dynamics that GaN nanowires with diameter of $6.44$~nm and length of $10$~nm has a thermal conductivity equal to $48$~W/mK.
We notice that the thermal conductivity of the GaN nanowires decreases by a factor of $65~\%$ and this reduction is much less important than in the case of Si or Ge nanowires, as mentioned in the introduction.  

\subsection{Thermal conductivity of GaN nanowires with edge and screw dislocations}

\begin{figure*}
	\begin{center}
		\includegraphics[width=10cm]{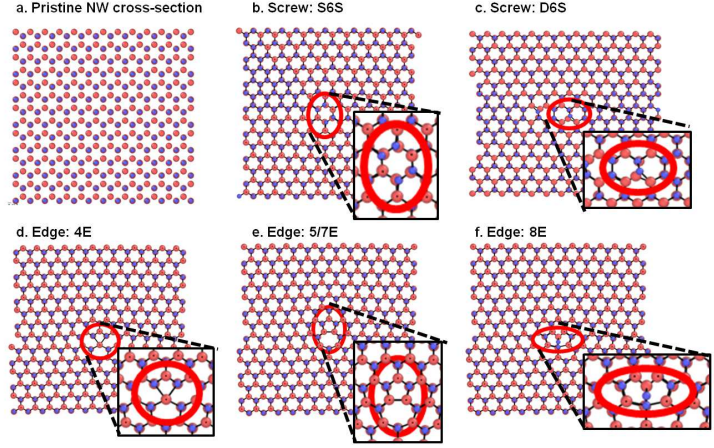}
		\caption{Cross-sections of the a. pristine nanowires with square cross-section, nanowires with screw dislocation which involves six atoms $S6S$ (b) and screw dislocation which involves twelve atoms $D6S$ (c). In d, e, f subfigures, there are nanowires with three types of edge dislocations, $4E$, $5/7E$, $8E$ respectively. With red cycles we marked the atoms involved directly to the dislocation.}
		\label{fig_3}
	\end{center}
\end{figure*}

\begin{figure}[h!]
	\begin{center}
		\includegraphics[width=8cm]{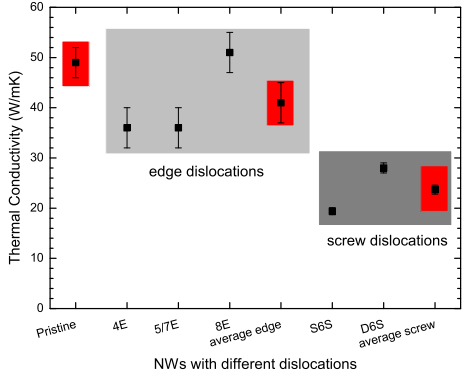}
		\caption{Thermal conductivity for pristine nanowires and nanowires containing difference types of dislocations. There are three sub-types of edge dislocations and two sub-types of screw dislocations. The light grey rectangular contains the edge dislocations and the dark grey the screw dislocations. The red rectangles inside the grey regions correspond to the average thermal conductivities of each type of defected nanowire.}
		\label{fig_4}
	\end{center}
\end{figure}

Nanowires with three different configurations of edge dislocations and two with screw dislocations, (figure~\ref{fig_3}) are modeled. The thermal conductivity in all nanowires with all five dislocations' configurations decreases compared to the pristine nanowire. The results are gathered to distinguish the influence of the two types of dislocations (figure~\ref{fig_4}). The discrepancy between nanowires with dislocations of different core configurations is quite remarkable. Let's concentrate first on the results concerning nanowires with edge dislocations. The thermal conductivity of the nanowire with the $8E$ dislocation is almost the same as the pristine nanowire, while the two other core configurations of edge dislocation have much lower thermal conductivities. The difference between the several core configurations is the nature of the bonds involved by the atoms at the core of the dislocations. In the case of $8E$ dislocation, there is a row of low coordinated atoms that undergo a dimerization along the dislocation line direction instead of exhibiting an sp3 hybridization that characterizes the bulk wurtzite structure\cite{Blumenau2000}. Hence, the atomic core configuration of the $8E$ dislocation exhibits low stresses and bonding conditions fairly similar to the perfect structure. On the other hand, in the nanowire with $5/7E$ dislocation, there is ``wrong" bonds established between atoms of the same species (Ga-Ga and N-N). These are always covalent bonds but with lower binding energy, meaning less strong in comparison with the Ga-N bonds at equilibrium. The lowest thermal conductivity appears for the nanowires with the $4E$ dislocation $38~\pm~4$~W/mK. In the latter nanowire, there is a row of over-coordinated atoms, leading to high stresses\cite{kioseoglou13}. Over-coordinated covalent bonds are more strained but less stiffer than the ``wrong" bonds and this reduces further the thermal conductivity. The same analysis is valid also for the nanowires with screw dislocations. Indeed, in the nanowire with the screw $D6S$ dislocation there are ``wrong bonds", while in the nanowires with the $S6S$ dislocation all bonds are between Ga and N atoms, associated with higher stresses which decrease further the thermal conductivity. The aforementioned analysis, gives a compact explanation of the impact of the different core configurations of dislocations on the thermal conductivity.

The issue to address now, is explaining the difference between the two types of dislocations in
impacting thermal conductivity (light grey square for the edge dislocations and dark grey square for the screw dislocations, figure~\ref{fig_4}). The reduction of the thermal transport due to phonons is more pronounced in the case of the nanowires with the screw dislocations $52~\%$ with an average thermal conductivity of $23.7~\pm~0.9$~W/mK, while the reduction of the edge dislocations is less than $20~\%$ compared to the pristine nanowires, with an average thermal conductivity of $41~\pm~4$~W/mK. As it will exposed in the following paragraph, the different impact that have edge or screw dislocations on the thermal conductivity of the nanowires can be related to their different associated elastic energies.

To understand the relative reduction of the thermal conductivity between the two types of dislocations, we analyzed some mechanical properties of the two kinds of defected nanowires. The calculated Young modulus for the bulk GaN in the [0001] direction found to be $365$~GPa and  $350$~GPa in the [100] direction, which is in a good agreement with experimental results\cite{Fujikane08,Yamaguchi99}. The difference of the Young modulus between the nanowires with the edge dislocation for the three configurations is $120$~GPa and for the screw dislocation $115$~GPa for the $S6S$ and $120$~GPa for the $D6S$. The values are almost the same for the two families of dislocations, and this clearly cannot explain the discrepancy of the calculated thermal conductivity between the two nanowires. The ratio between the average thermal conductivity of the nanowires with screw dislocations and the edge dislocation is $\frac{TC_{screw}}{TC_{edge}} = \frac{24}{44}=0.54$ and this reduction by a factor of two cannot be assigned to their Young moduli.

\subsection{Elasticity theory}

In order to understand better the difference of the thermal conductivity between the two types of dislocations, we have calculated analytically the strain energy of an infinite straight dislocation in a perfect crystal using linear elasticity theory\cite{hirth82}. The total strain energy of a dislocation is equal to the sum of elastic and core energies: 

\begin{equation}
E_{total}=E_{elastic}+E_{core}
\end{equation} 

The elastic energy per unit length of a dislocation ($r_0 \leq R$) contained in a cylinder of radius R is given by

\begin{equation}
E_{elastic}= Aln(\frac{R}{r_0})
\label{eq_6}
\end{equation} 

where  

\begin{equation}
A = \frac{K \textbf{b}^2}{2\pi}
\end{equation} 

$A$ is the prelogarithmic factor, \textbf{b} is the magnitude of the Burgers vector, $K$ is an energy factor and $r_0$ is the core radius. 

The energy factors for the screw ($K_s$) and edge ($K_e$) perfect dislocations following the anisotropic elasticity theory for a hexagonal structure [1] are equal to:

\begin{equation}
K_s = C_{44}
\end{equation} 

\begin{equation}
K_e = \frac{C_{11}^2-C_{12}^2}{2C_{11}} 
\end{equation} 

where $C_{11}$, $C_{12}$, and $C_{44}$ are elastic constants.
Using the elasticity theory, the prelogarithmic factors for the edge and the screw dislocations were calculated, as well as the ratio between them. Using elastic constants calculated by the MSWp\cite{aichoune00}, the prelogarithmic factors are found to be $A_e$=0.77~eV/\AA ~and $A_s$=1.46~eV/\AA ~leading to a ratio equal to $\frac{A_e}{A_s}$=0.53. If experimental determined elastic constants\cite{polian96} are used the prelogarithmic factors are $A_e$=0.85~eV/\AA ~and $A_s$=1.38~eV/\AA , and the ratio between them equals $\frac{A_e}{A_s}$=0.62. 
The elastic energy of a dislocation is related to the radius of the cylinder and it is infinite for an infinite crystal. However, it can be calculated for a finite radius, and the calculated thermal conductivity is found inversely proportional to the stored elastic energy for the two perfect threading dislocations.

It is well known that there are big differences on the stress fields between the edge and the screw dislocations in wurtzite GaN. For the screw dislocation the induced stress on the plane perpendicular to the dislocation line is pure shear stress with complete radial symmetry, while for the edge dislocations there are tensile and compressive lobes. These characteristics may explain the differences identified in thermal conductivity in figure~\ref{fig_4} between the edge and the screw threading dislocations. Moreover, the high elastic strain energy of the screw dislocation compared to that of the edge one is due to the difference between the magnitudes of their Burgers vectors. Then, a screw dislocation induces more strained Ga-N bonds in the elastic region than does an edge dislocation. This may explain why a screw dislocation reduces the thermal conductivity of GaN than the edge dislocation.

It should be noticed that the important impact of the screw dislocation on thermal conductivity is analogous to their impact on optoelectronic properties. As it is proved by J. Abell and T.D. Moustakas~\citep{Abell2008}, the suppression of the nonradiative recombination in screw dislocations leads to an increment by a factor of 2 of the internal quantum efficiency of InGaN multiple quantum wells grown on GaN. This leads to the conclusion that although the majority of the threading dislocations in GaN are of the edge type, the most crucial dislocations either for the optoelectronic or the thermal properties of the devices are the screw dislocations. As it is elucidated in the current study, the edge dislocation does not impact severely the thermal properties of the GaN. Moreover, following the experimental results of Ref~\citep{Abell2008}, this is also valid for the their optoelectronic properties, and hence, their overall influence on the GaN based device performance should not be considered detrimental.

\begin{figure*}
	(a)\includegraphics[width=0.5\linewidth]{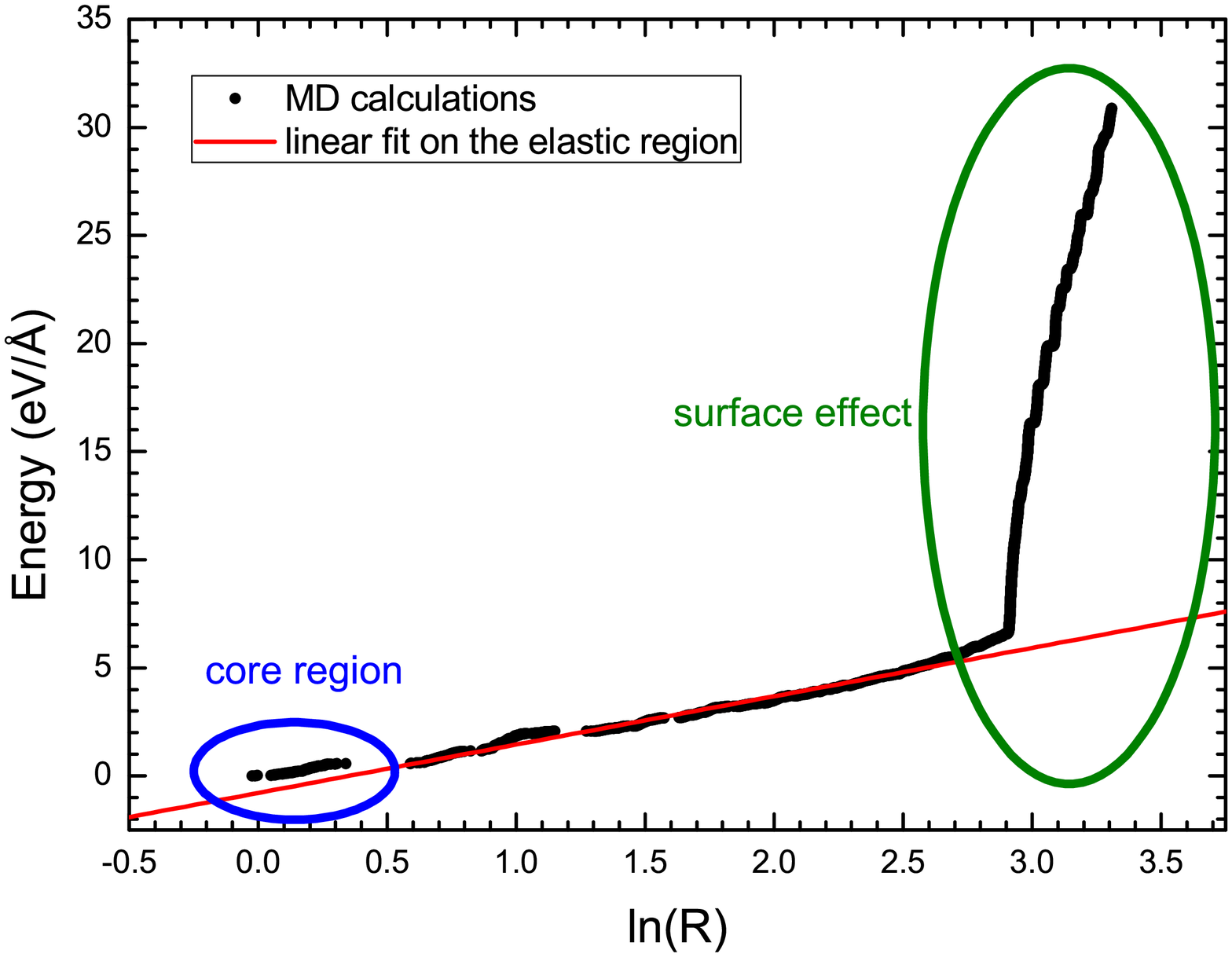}
	(b)\includegraphics[width=0.5\linewidth]{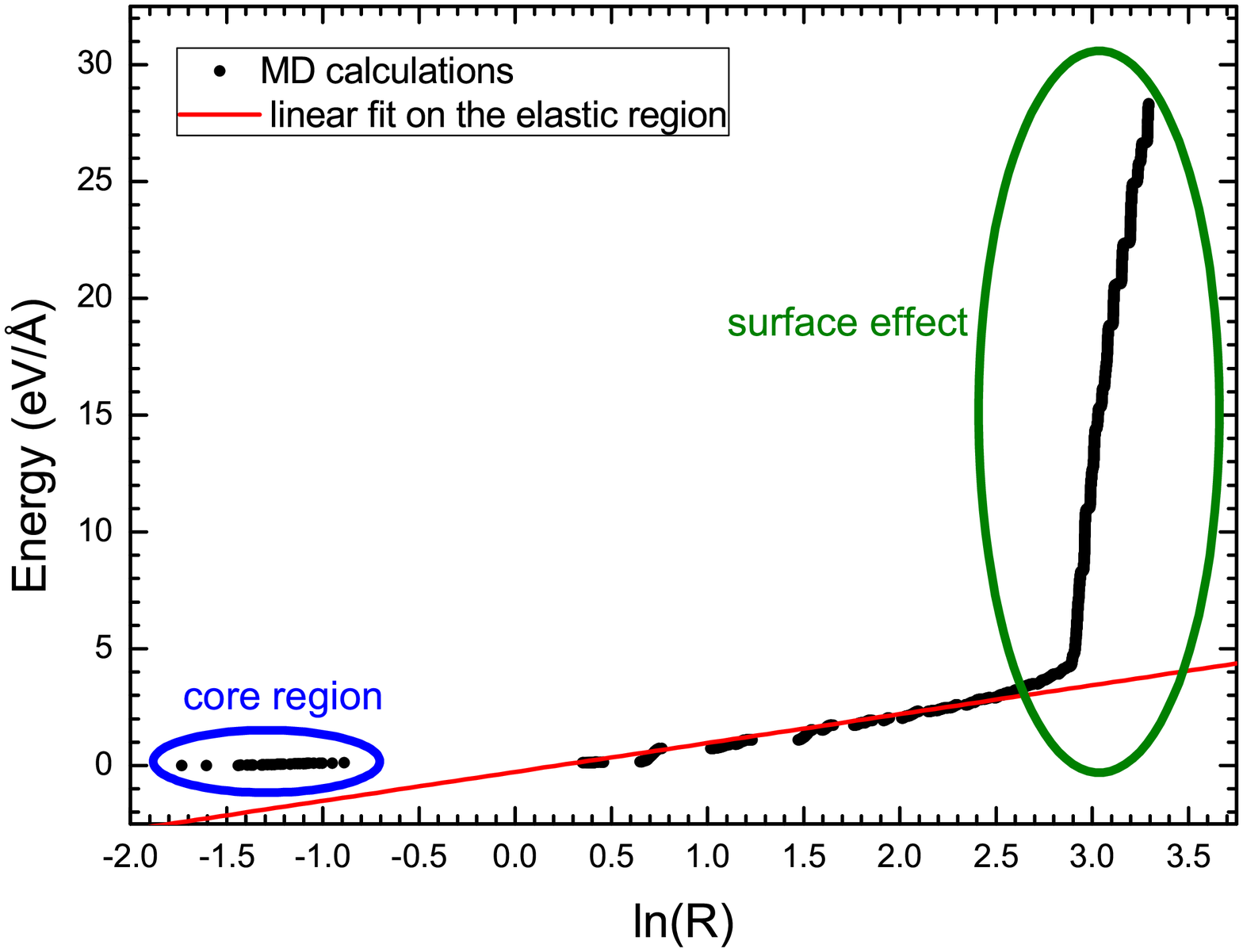}
	\caption{(Color online) Energy stored in a cylinder of radius $R$ as a function of $ln(R)$ for the (a) $S6S$ and (b) $4E$ dislocations. The three identified regions are denoted. The core region, the elastic region where the linear fit is performed in order to evaluate the prelogarithmic factor A, and the surface region.}
	\label{fig_new}
\end{figure*}

In order to evaluate the elastic energy of the dislocations, the energy in the region bounded by a cylinder of radius  $R$ is plotted versus $ln(R)$. The core radius $r_0$ is taken at the point where the curve stops being linear $r_0 \leq R$ and consequently Eq. 6 can be written as:

\begin{equation}
E_{elstic}(r) = A lnR - Alnr_0
\end{equation}

The corresponding prelogarithmic factor is evaluated by the above fitting equation to the calculated by MD values. 
Following the procedure described above, figure~\ref{fig_new} illustrate the energy versus $ln(R)$ plots for the (a) $S6S$ and (b) $4E$ dislocations. In both cases, it can be observed three regions. Initially the core region where the atoms are bonded following the structural characteristics of the simulated core configuration, after that the curves become linear and the energy adopts the expression given by elasticity theory. After the elastic region, the energy is suddenly increased due to the surface effects of the considered configurations. Due to the surface relaxation of the considered models the elastic region is influenced in such a way that the slope of the linear elastic region is increased with respect to the values expected by the elasticity theory. This behavior that the presence of free surfaces leads to a quick enhancement of the strain energy has been also identified previously~\cite{Belabbas2007,Belabbas2013b}. The slope of the elastic region, following the above equation, corresponds to the prelogarithmic factor $A$. The prelogarithmic factors are found to be $A_s$=2.24~eV/\AA ~and $A_e$=1.24~eV/\AA ~leading to a ratio equal to $A_e/A_s$=0.55 in agreement with the elasticity theory prediction $A_e/A_s$=0.53. It is also evaluated the spatial extension of the surface effect which is found equal $\sim$~9~\AA ~for both screw and edge dislocations.

\subsection{Temperature dependence of the thermal conductivity} 

\begin{figure*}
	(a)\includegraphics[width=0.5\linewidth]{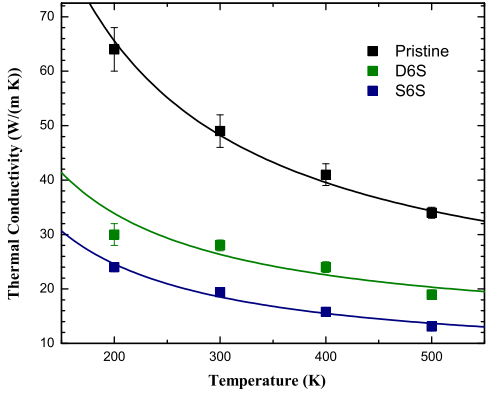}
	(b)\includegraphics[width=0.5\linewidth]{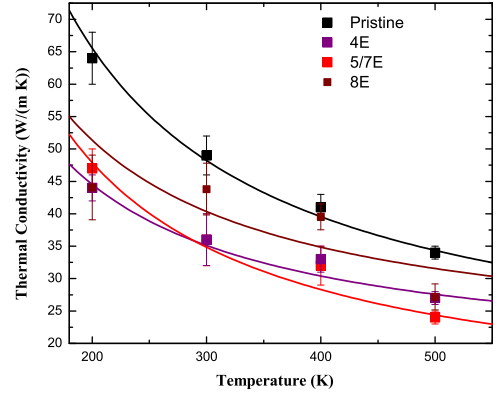}
	\caption{(Color online) Thermal conductivity of the pristine and the two screw (a) and three edge (b) defected nanowires as the function of the temperature. The solid lines represent the above described theoretical approach.}
	\label{fig_7}
\end{figure*}

The temperature dependence of the thermal conductivity of pristine nanowires and those containing two types of dislocations were appraised to demonstrate the existence or not of anharmonic effects. Generally, the thermal conductivity of all types of nanowires is decreasing by the enhancement of the temperature. This was already demonstrated in GaN layers by Kotchetkov et al~\cite{kotchetkov01} who combined both experimental data with an analytical model. In figure~\ref{fig_7}(a and b) one can observe that for pristine nanowires the reduction of the thermal conductivity as a function of the temperature is much faster than in the nanowires containing dislocations. The different types of dislocations reduce in a different way the thermal conductivity and thus affect in a different manner its temperature dependence ~\cite{Kamatagi2009}. 

For the nanowire with the $5/7E$ dislocation the rate of the thermal conductivity reduction is less
but still similar to that of the pristine nanowire. Otherwise, the slope is getting less and less
important in the $4E$, $D6S$ and $S6S$ nanowires. This could be attributed to the fact that the
anharmonic effects become less important in last three nanowires than for the pristine nanowire
and that with the $5/7E$ dislocation. This might be related to the fact that in the case of the
defected nanowires, the normal scattering and the phonon-dislocation scattering continues to play a
predominant role, while in the pristine and nanowire with the $5/7E$ dislocation, the Umklapp
process dominates, resulting in a rapid reduction of the slop. In the case of the $8E$
configuration the picture is different as the core contains dangling bonds. We then observe an
important reduction of the thermal conductivity between 400~K and 500~K, while it is almost constant at lower temperatures. It is interesting to note here that Kamatagi et
al~\cite{Kamatagi2007} have shown that an increase in the dislocation scattering strength shifts the
temperature of the maximum thermal conductivity towards higher temperatures. This might explain
the different temperature dependences of the thermal conductivity of $8E$ and $D6S$ nanowires.

We observe also a crossing of the thermal conductivity curves of the
nanowires with $5/7E$ and $4E$ dislocations. The values are close but we can observe a tendency
according to nanowires with dislocations which contain ``wrong bonds" have more severe impact on
the thermal conductivity in high temperatures than the usual Ga-N covalent bonds. The different
behavior of phonons within structures with dislocations might be related with different spatially
localized modes associate with the different types of dislocations~\cite{kang14}. It is not on the
scope of the current study, but it will be interesting to investigate the dynamic properties of defects
as the vibrational lifetimes of phonons trapped in each kind of
dislocations~\cite{gibbons11,estreicher14,kang14,estreicher15}.         

It is well known that the effect of uniform elastic strain on thermal conductivity can be elaborated by means of Peierls~\cite{Peierls2001} formulation as it is presented by Somnath Bhowmick and Vijay B. Shenoy~\cite{Bhowmick2006}. Taking into account that the phonon frequencies are expected to scale with strain as $\omega = \omega_0 \varepsilon^{-\alpha}$, 
where $\varepsilon$ is the uniform elastic strain and $\varepsilon < 1 $ for compression while $\varepsilon > 1 $ for tension, and $\alpha$ is a non-universal (potential dependent) parameter, it is concluded that the thermal conductivity $\kappa$ is analogous to $T^{-1} \varepsilon^{\gamma}$, where $\gamma$ is a material dependent parameter~\cite{Bhowmick2006}. This dependence of the
thermal conductivity has been validated previously by MD simulations~\cite{Tretiakov2004} and has
been extensively used in a variety of bulk and nanostructured materials~\cite{Li2010,Parrish2014} in
order to elucidate the elastic strain effect on thermal properties.
The crystal lattice is heavily and not uniformly strained around a
dislocation. Consequently, the physical properties influencing the phonon dispersion in the
crystalline lattice, such as the mass density and the forces between the neighboring atoms,
considerably deviate in the dislocated area from other parts of the crystal that are under uniform
elastic strain. Definitely, the scattering of phonons on the disturbed areas due to the dislocation
stress fields, reduces the lattice thermal conductivity. The
dislocation stress and consequently strain field of edge and screw dislocations exhibit a 3D
dispersion, which is quite different from the uniform elastic strain. Only the elastic energy (Eq.~\ref{eq_6}) presents a linear behavior with the logarithm of the radius from the
core of the dislocation in a coaxial cylinder with the dislocation line. It is well
known that the elastic energy is proportional to the elastic strain under the limits of the elasticity
theory and hence, the phonon frequencies are anticipated to vary with elastic energy. Consequently
the thermal conductivity of dislocations should follow the relationship:

\begin{equation}
\kappa \propto T^{-1} E_{elastic}^J
\end{equation}

where $J$ is a material and more specifically dislocation’s core dependent parameter. 
If we consider Eq. 6, the thermal conductivity should be related with the prelogarithmic factor of the dislocations in the following way:

\begin{equation}
\kappa \propto T^{-1} A^J
\label{eq_new}
\end{equation}

For the pristine structure it should be considered just $\kappa \propto T^{-1}$ for the temperature range under investigation (Umklap phonon-phonon interaction). 
The thermal conductivity data calculated by the MD simulations are fitted to Eq.~\ref{eq_new}, taking into account the aforementioned calculated prelogarithmic factors $A_s$=2.24~eV/\AA ~and $A_e$=1.24~eV/\AA ~and the $J$ factor for each investigated dislocation as well as the corresponding core coordination characteristics are given in Table~\ref{table1}. The $J$ factor values, that are calculated, lead to the following conclusion. The structural characteristics of the core and more specifically the coordination which characterize each individual core structure rule the corresponding $J$ factor. In particular, the highest $J$ factor corresponds to the core configuration which is closer to the pristine structure, and the $J$ factor is decreased when the core coordination is diversified. For both screw and edge dislocations the lowest $J$ factor is calculated for the ``wrong" bonded core configurations. It should be noticed here that, the energetical impact of over coordination is lower that the under coordination by the use of the MSWp\cite{bere02}. Hence, the $4E$ edge dislocation is closer to the perfect GaN than the $8E$ core configurationand in many cases $4E$ edge dislocation has been denoted as perfect coordinated core configuration\cite{Lymperakis2004}. Moreover, it is expected and proved with the above arguments that screw dislocations having higher strain energy, i.e. more strained bonds, hinder the thermal transport more than edge dislocations, which are associated with lower elastic energy.

\begin{table}[b]
	\centering
	\caption{The $J$ factor and the corresponding core coordination characteristics for each investigated dislocation.}
	\begin{tabular}[htbp]{@{}lccc@{}}
		\centering
		\textbf{Type of }     &  & \textbf{J}        & \textbf{Core coordination} \\ 
		\textbf{Dislocation}   &    & \textbf{factor}        & \textbf{characteristics} \\
		\textbf{S6S}              &         & 1.3                      & normal coordinated                        \\
		\textbf{D6S}               &        & 1.0                      & wrong coordinated                          \\
		\textbf{4E}                &        & 2.8                      & over coordinated                           \\
		\textbf{8E}                &        & 2.1                      & low coordinated                            \\
		\textbf{5/7E}              &        & 1.3                      & wrong coordinated                          \\ 
	\end{tabular}
	\label{table1}
\end{table}

\subsection{Dislocation density dependence of the thermal conductivity} 

\begin{figure}[h!]
	\begin{center}
		\includegraphics[width=8cm]{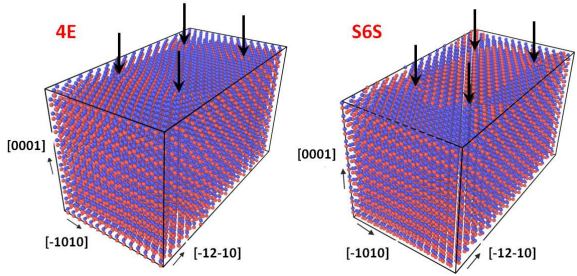}
		\caption{Defected bulk GaN with density of dislocations 94.5 10$^{11}$cm$^{-2}$ for (a) edge (4E) and (b) screw (S6S) configurations at 300K. The black arrows show the cores of the dislocations.}
		\label{fig_8}
	\end{center}
\end{figure}

It has been observed experimentally~\cite{Kogure1975,Mion2016b,Mion2006} and demonstrated theoretically~\cite{kotchetkov01,Zou2002} that there are two regimes of the thermal conductivity dependence over the density of dislocations: the low and high dislocation density regimes. In low density regime, the thermal conductivity does not vary upon the dislocations' density, while in high density regime the thermal conductivity varies as the inverse of the logarithmic of the dislocation density. In experiment the threshold is at $10^{7}$~cm$^{-2}$,~\cite{Mion2016b,Mion2006} while in simulations is at $10^{11}$~cm$^{-2}$.~\cite{kotchetkov01,Zou2002} Our modeling with MD is for dislocations' densities equal and greater than $10^{12}$~cm$^{-2}$, thus we are in high dislocation regime. Two bulk GaN systems with edge and screw dislocation with density of dislocations 94.5 10$^{11}$cm$^{-2}$  are depicted in figure~\ref{fig_8}, left and right respectively. The black arrows indicate the positions of the cores of the direction of the dislocation lines.

\begin{figure*}
	(a)\includegraphics[width=0.5\linewidth]{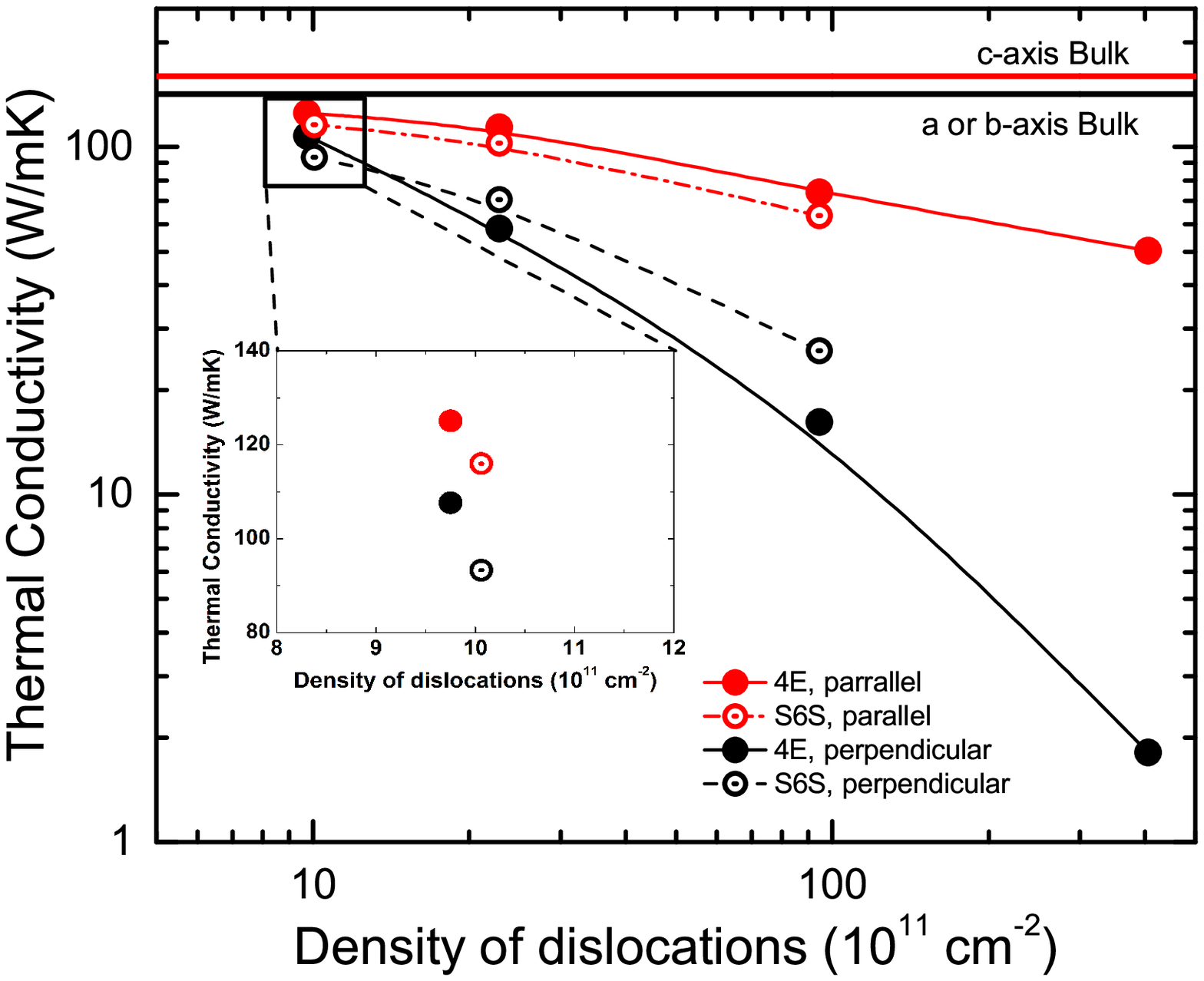}
	(b)\includegraphics[width=0.5\linewidth]{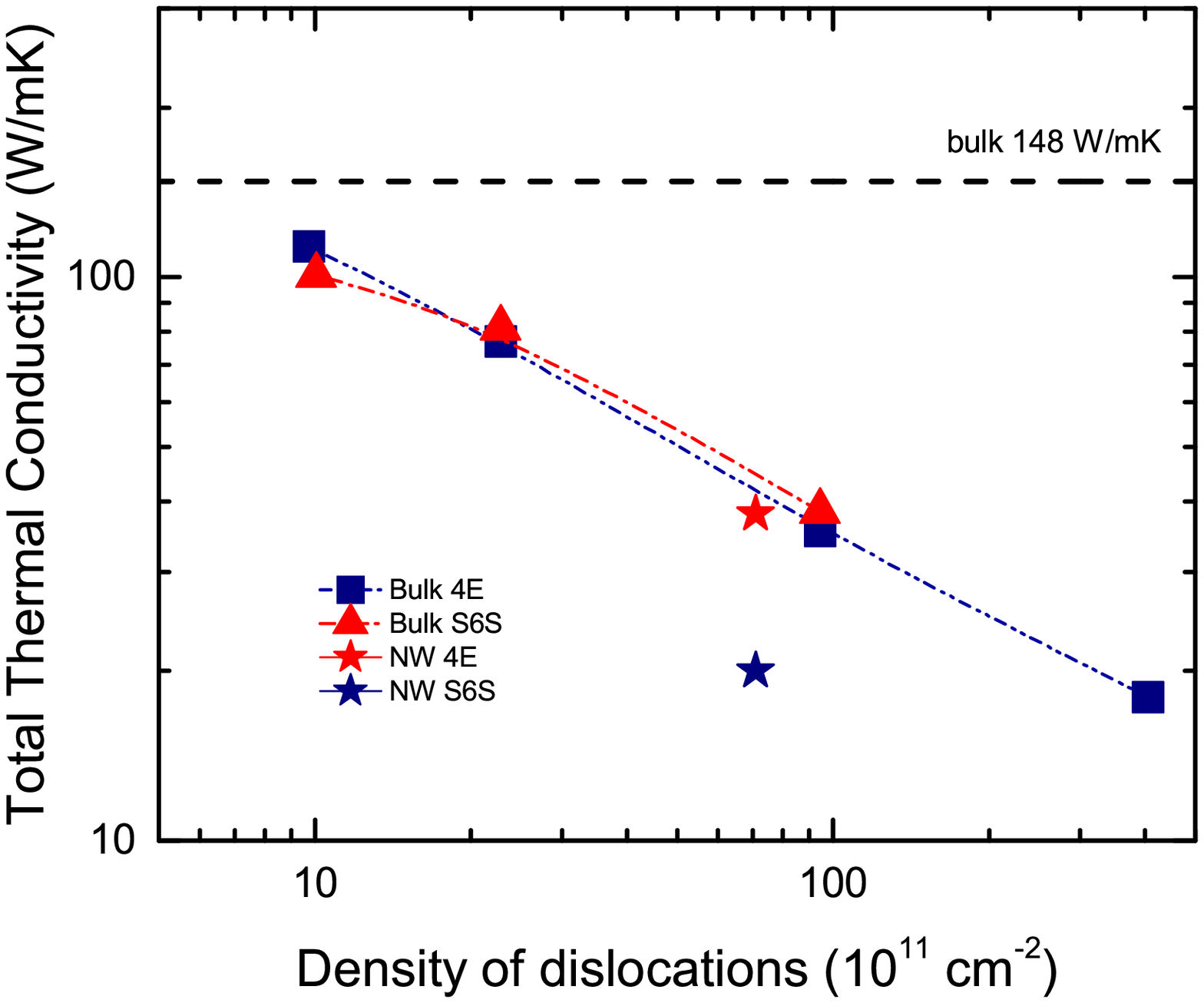}
	\caption{(Color online) Thermal conductivity of bulk GaN as a function of dislocations' density for the edge $E4$ and the screw $S6S$ configurations (a) thermal conductivity tensors components parallel and perpendicular to the dislocation lines (b) total thermal conductivity for bulk and nanowires with dislocations. The straight lines of the total and the a,c axis thermal conductivity vectors of the crystalline with no defects bulk GaN are added to help the discussions.}
	\label{fig_9}
\end{figure*}

The thermal conductivity tensor components, parallel and perpendicular to the dislocation lines, are depicted in figure~\ref{fig_9}-(a) for bulk GaN as a function of the screw $S6S$ and edge $4E$ dislocation density. First we notice that the evolution of the thermal conductivity parallel and perpendicular to the dislocation lines as a function of the density of dislocations is quite different. The thermal conductivity tensor component parallel to the dislocations is much less effected than the vector perpendicular to the dislocations when varying the density of dislocations. This means that phonons are much less perturbed by dislocations when they propagate in the same direction as the dislocation's line. The thermal anisotropy is enhanced by increasing the dislocation density. We then pass from anisotropy of $1.2$ for the edge and $1.3$ for the screw for a density of 10$^{12}$cm$^{-2}$ to $4.6$ for edge and $3.3$ for screw for a density of 10$^{12}$cm$^{-2}$. Interestingly, there is no discrepancy on the variation of the total thermal conductivity of the bulk GaN as a function of the dislocation density for the systems with the two types of dislocations (figure~\ref{fig_9}, right). Here we observe the logarithmic dependence of the total thermal conductivity on the density similar to what it has been reported before~\cite{Mion2016b,Mion2006,kotchetkov01,Zou2002}. The thermal conductivity decreases rapidly by increasing the dislocation density. Our results are in good agreement with previous theoretical results~\cite{kotchetkov01}.

To compare the impact of the dislocation density on bulk system and nanowires, we added in figure~\ref{fig_9}-(b) the thermal conductivity of the corresponding defected nanowires. The $4E$ dislocation has the same impact on the thermal conductivity of nanowires and bulk. In contrast the $S6S$ screw dislocation degrades the thermal conductivity of nanowires in much more severe way than in bulk GaN. This is owing to the stronger coupling between the screw dislocations and the surfaces of the nanowire than for the edge dislocations.

\section{Conclusions}

The interaction between phonons and the core of dislocations and their strain field leads to the decrease of the thermal conductivity. The dislocations that distort the lattice in large spatial region as the screw dislocations scatter effectively the phonons and especially those with large wavelengths\cite{gruner68}. 

Edge and screw dislocations have different impact on the thermal properties of GaN nanowires. The difference might be quite important depending on the type of the dislocation and the core configuration. The nature of the bonds involved by the atoms at the core of the dislocation explains qualitatively the impact of each type of dislocation. Ga-N covalent bonds around the dislocation are stiffer and reduce the thermal transport, while the Ga-Ga or N-N ``wrong bonds", which are ``softer", influence less the phonon transport. Finally dangling free bonds seems to have minimal influence on the thermal transport. Screw dislocations reduce the thermal conductivity by a factor of two compared to the edge dislocations. The physical origin of the effect of dislocations has been explained in the framework of linear elasticity theory and MD simulations' results, with which we found that the thermal conductivity is inverse proportional to the elastic energy of the dislocations. The temperature dependence of the thermal conductivity varies also depending on the type of dislocation. A physical law that elucidates the temperature dependence of the TC of isolated dislocations with an exponent factor which depends on the material, dislocation's type and the core characteristics has been described. The effect of the strain field of the screw dislocations is found to be more pronounced than this of the edge dislocations.


Concerning the impact of the density of dislocations on the thermal conductivity of bulk GaN, we confirmed previous results: high dislocation density reduces rapidly the total thermal conductivity following a logarithmic dependence. A striking point is that the thermal conductivity tensor components parallel and perpendicular to the dislocations lines show different dependence on the dislocation density. Furthermore the anisotropy of the thermal tensors grows in increasing the density of dislocations. Comparing the results of the impact of dislocations on the thermal properties of nanowires and bulk system, we revealed a different behavior of the screw dislocations in decreasing the dimensions of the system. It seems that screw dislocations interacts stronger with the surface of nanowires than edge dislocations, decreasing further the phonon transport.

\section*{Acknowledgments}

The authors are grateful for the use of the large computer facilities of 'ARIS' National HPC Infrastructure of the Greek Research and Technology Network and of 'ERMIONE' cluster (IJL-LEMTA) and IDRIS.

\bibliographystyle{unsrt}
\bibliography{biblio_GaN_PRB}

\end{document}